\documentclass[grl]{agu2001}
\usepackage{graphicx}
\authorrunninghead{PIEGARI et al.}

\titlerunninghead{A cellular automaton for landslides}

\authoraddr{
piegari@na.infn.it,
cataudella@na.infn.it, dimaio@na.infn.it, milano@na.infn.it,
mario.nicodemi@na.infn.it}

\begin{document}

%
%
%
%
%

%
%

\title{A cellular automaton for the factor of safety field in landslides modeling}

%

%
%


\author{E. Piegari, V. Cataudella, R. Di Maio, L. Milano, and M. Nicodemi}
\affil{Dip. di Scienze Fisiche, Universit\'a di Napoli ``Federico II'',
INFN, CRdC AMRA, CNR-Coherentia, Napoli, Italy}




%
%


\begin{abstract}
Landslide inventories show that the statistical distribution of the area of
recorded events is well described by a power law over a range of decades.
To understand these distributions, we consider a cellular automaton to model
a time and position dependent factor of safety. The model is able to
reproduce the complex structure of landslide distribution, as
experimentally reported. In particular, we investigate the role of
the rate of change of the system dynamical variables, induced by an
external drive, on landslide modeling and its implications on hazard
assessment. As the rate is increased, the model has a crossover from
a critical regime with power-laws to non power-law behaviors. We
suggest that the detection of patterns of correlated domains in
monitored regions can be crucial to identify the response of the
system to perturbations, i.e., for hazard assessment.
\end{abstract}

%
%

%

\begin{article}

%
%

{\bf Introduction } As for earthquakes and forest fires, there is a
compelling evidence that the landslide frequency-size distributions
are power-law functions of the area \citep{Turcotte_02}. The
presence of these broad distributions has crucial consequences on
both the basic understanding of these phenomena and the practical
and relevant purposes, such as the evaluation of natural hazards.
Here, we introduce a cellular automaton that is aimed at modeling
the general features of landslides, and is focused on the dynamical
evolution of a space and time dependent factor of safety field. This
model is very simple, but it is able to give a comprehensive picture
of the avalanching phenomena and to reproduce some well-known
properties of landslide distributions.

Several authors invoked the paradigm of self-organized criticality (SOC)
\citep{Bak_87,Jensen_98,Turcotte_99}
to explain landslide distributions
\citep{Turcotte_02,Pelletier_97,Hertgarten_00}.
Although the ``critical'' nature of the present phenomenon is not
yet assessed and many authors believe that deviations from power-law
appear to be systematic for small landslides data
\citep{Stark_01,Pelletier_97,Brardinoni_04,Malamud_04},  several
regional landslide inventories records show robust power-law
distributions of large events with an exponent around $\alpha \sim
2.5$ \citep{Turcotte_02}, ranging approximately from $\alpha \sim
1.75$ for rockfalls to $\alpha \sim 2.8$ for mixed landslides (see
\citep{Dussauge_03,Faillettaz_04} and references therein). The
universality of such an exponent is still debated (see
\citep{Turcotte_02,Dussauge_03,Faillettaz_04,Malamud_04} and
references above), and its reported values are far from the one in
the original ``sandpile model''  \citep{Bak_87}, where $\alpha \sim
1.0$. Recently, the reported values of $\alpha$ have been obtained
by introducing two-thresholds mechanisms in models that relate
landslide dynamics both to SOC \citep{Pelletier_97,Hertgarten_00}
and to non SOC cellular automata \citep{Faillettaz_04}. Furthermore,
models based on $\Gamma$ \citep{Malamud_04} or Pareto
\citep{Stark_01} distributions have been proposed.

Here, we consider a model inspired to an {\em anisotropic} version
of the Olami-Feder-Christensen (OFC) \citep{Olami_92} cellular
automaton and subject to a {\em finite driving rate}
\citep{Hamon_02}. The model describes the evolution of a space and
time dependent factor of safety, which is investigated for the first
time in the present framework. In particular, we outline the
essential role played by the rate of change of the system dynamical
variables (variation of pore water pressure, lithostatic stress,
cohesion coefficients, etc., \citep{Helley_04,Iverson_00}) induced
by external triggers. We find the model to be at the edge of the SOC
limit. Actually, such a limit, which is achieved only in the
asymptotic condition of vanishing driving rate, is hardly attainable
in a real landslide process. The model is able to reproduce
power-law distributions with exponents very close to the observed
values. Power-laws are robust even though their exponent smoothly
depends on system parameters (e.g., time derivative of the factor of
safety and its dissipation level, see below). In this sense,
although the SOC paradigm to some extent may be applied to
landslides \citep{Turcotte_02}, the idea of universality, within
this model, must be restricted to the shape of the frequency-size
distribution rather than to its exponent, as can be deduced from
some catalogues \citep{Dussauge_03,Faillettaz_04}. Finally, in
presence of strong driving rates we find that the model has Gaussian
behaviors. We examine below the implications of our results on
hazard assessment.

{\bf The Model } The empirical Mohr-Coulomb failure criterion
establishes that landslides occur when the shear stress exceeds a
maximum value, which is given by $\tau_{max}=c + (\sigma
-u)\tan\phi$, with $\sigma$ the total normal stress, $u$ the
pore-fluid pressure, $\phi$ the angle of internal friction of the
soil and $c$ the cohesional (non-frictional) component of the soil
strength \citep{Terzaghi_62}. In literature, the factor of safety,
$FS$, against slip is defined by the ratio of the maximum shear
strength $\tau_{max}$ to the disturbing shear stress $\tau$
\begin{equation}
FS = \frac{\tau_{max}}{\tau}.
\end{equation}
If $FS>1$ resisting forces exceed driving forces and the slope
remains stable. Slope failure starts when $FS = 1$. Although the
practical determination of $FS$ is difficult, simple one-dimensional
infinite-slope models can quantify how $c,u$ and $\phi$ influence
the Coulomb failure and show that the ground-water term has the most
widely ranging influence \citep{Iverson_97}. Traditional models
generally treat soils and rocks as continuous porous media that obey
to the Darcy's law. Actually, field evidence indicates that the
hydrology of some natural slopes is strongly influenced by
discontinuities such as fractures and macropores. In practice,
observations of large spatial and temporal fluctuations of water
flow, within slopes at different sites, support the assertion that
water-flow paths and permeability continually change within the
slopes and also during the failure, providing different local values
of the pore pressures and of the cohesion
\citep{Iverson_97,Helley_04,Iverson_00}.

In order to take into account the complex non-homogeneous structure
of a slope in the above failure condition, we consider a site and
time dependent factor of safety, $FS$. In particular, we approximate
a natural slope by a two-dimensional (square) grid and define on
each cell, $i$, of the lattice a local variable $e_i = 1/FS_i$. Such
a local inverse factor of safety is the fundamental dynamical
variable of our model. The presence of diffusion, dissipative and
driving mechanisms acting in the soil, such as those on the water
content, inspires the dynamics of our model, which is defined by the
following operations. Starting from a random and ``stable'' initial
configuration ($e_i<1$ $\forall i$), the system is subject to
changes caused by some external trigger, as for instance a uniform
rainfall, and the values of $e_i$ on each cell of our grid change at
a given rate $\nu$, $e_i \rightarrow e_i + \nu$. For the sake of
simplicity, we consider here only a uniform driving rate, $\nu$, but
different choices can be made to simulate the effect of different
hydrologic and external triggering mechanisms. The model is driven
as long as $e_i < 1$ on all sites $i$. Then, when the generic site
$i$ becomes unstable (i.e., overpasses the threshold, $e_i\geq 1$),
it relaxes with its neighbors according to the rule:
\begin{itemize}
\item[] $e_i \rightarrow 0$; \quad \quad $e_{nn} \rightarrow e_{nn}+f_{nn} e_i$,
\end{itemize}
\noindent where the index $nn$ denotes the nearest neighbors of site
$i$ and $f_{nn}$ is the fraction of $e_i$ toppling on $nn$ (after
failure we set $1/FS=0$ for simplicity, as any other finite level
would work \citep{Jensen_98}). This kind of chain relaxations
(``avalanches'') is considered to be instantaneous compared to the
time scale of the overall drive and it lasts until all sites are
below threshold. The model is said to be conservative if
$C=\sum_{nn} f_{nn} =1$. Since many complex dissipative phenomena
(such as evaporation mechanism, volume contractions, etc.
\citep{Fredlund_93}) contribute to a dissipative stress transfer, we
consider the non-conservative case $C<1$, which is different from
previous landslide models \citep{Pelletier_97,Hertgarten_00}. Since
gravity individuates a privileged direction, we consider an
anisotropic model where the fraction of $e_i$ moving from the site
$i$ to its ``downward'' (resp. ``upward'') neighbor on the square
lattice is $f_d$ (resp. $f_u$), as $f_l=f_r$ is the fraction to each
of its ``left'' and ``right'' neighbors. In particular, we assume
$f_u < f_d$ and $f_l =f_r < f_d$. This choice of parameters is made
in the attempt to sketch the complex relaxation processes occurring
in a slope failure. The conservation level, $C$, and the anisotropy
factors, $f$'s, which we assume to be uniform, are actually related
to the local soil properties (e.g., lithostatic, frictional and
cohesional properties), as well as to the local geometry of the
slope (e.g., its morphology). The rate of change of the inverse
factor of safety, $\nu$, which is induced by the external drive
(e.g., rainfall) and is related to soil and slope properties,
quantifies how the triggering mechanisms affect the time derivative
of the FS field.

{\bf Numerical Results }
We consider a $64 \times 64$ square
lattice, implementing both cylindrical (open along the vertical axis
and periodical along the horizontal axis) and open boundary
conditions, which do not give appreciable differences.
Once the system has attained a stationary state in its dynamics, we
study the probability distribution, $P(s)$, of avalanches of size
$s$. During a run (we treat statistics of over $10^9$ events per
run) the conservation level $C$ and the rate, $\nu$, are kept fixed.
Examples of the frequency-size distribution of avalanches, $P(s)$,
are shown in figures \ref{scaling} and \ref{scaling_C}. In figure
\ref{scaling}, the different curves correspond to different values
of the rate $\nu$, for $C=0.4$, and in figure \ref{scaling_C} to
different values of $C$, for $\nu=5\cdot 10^{-3}$.

In the limit of very small driving rate, i.e., $\nu\rightarrow 0$,
the distribution of events, $P(s)$, exhibits the typical SOC
structure (see figure \ref{scaling}): a power law characterized by a
critical exponent $\alpha$, $P(s)\sim s^{-\alpha}$, in agreement
with the experimental evidence on medium and large landslides
\citep{Turcotte_02,Pelletier_97,Hertgarten_00,Faillettaz_04,Brardinoni_04,Malamud_04,Dussauge_03},
followed by a size dependent exponential cutoff \citep{Jensen_98}.
By increasing the rate  $\nu$, the power-law regime shifts towards
larger sizes and at some point the probability distribution
apparently shows a maximum for a value $s^*$. There are two regimes:
for large landslides ($s>s^*$) the above structure $P(s)\sim
s^{-\alpha}$ is found, while for small events ($s<s^*$) an
increasing function of $s$ is observed. Such a complex structure
is absent in SOC models, instead a maximum is found in landslide
inventory maps for small landslide data, although there is no
consensus about the nature of such a feature
\citep{Stark_01,Brardinoni_04,Malamud_04}. The values of the
power-law exponent, $\alpha$, by varying the rate, $\nu$, and $C$
are very close to those experimentally found
\citep{Turcotte_02,Dussauge_03}. As in the original isotropic OFC
model \citep{Olami_92}, the critical exponent decreases with the
level of conservation, $C$ (see inset in figure \ref{scaling_C}).
The value of $\alpha$ slightly changes with the anisotropic ratios
$f_d/f_d$ and $f_u/f_d$, except when they get too small
\citep{Piegari_05} where, as found also in other models
\citep{Amitrano_99, Amitrano_03, Faillettaz_04}, the event size
distribution is considerably modified.
The power-law regime is crucially robust to changes in system
parameters. For instance, in the case of figure \ref{scaling} it can
be found for $\nu$ up to approximately $10^{-2}$, all over the range
$C\in[0.4,0.8]$. It is worth noting that the $\alpha$ values here
obtained are comparable to those found in models of failure in fiber
bundle \citep{Hansen_Hammer92,Hansen_Hammer94,Hidalgo_02}.

As it can be seen in figure \ref{scaling}, a further increase of the
driving rate (above $\nu\sim 10^{-2}$) causes a crossover to a
markedly different regime where power-laws are no longer apparent
and a bell shaped (Gaussian) distribution emerges, whose peak shifts
towards larger sizes and shrinks up. Such a behavior is to be
expected since for strong driving rates all internal correlations
are washed out.

Summarizing, the conservation level, $C$, and the time derivative of
$1/FS$, $\nu$, turn out to be important to determine landslide
probability distributions: in the limit $\nu\rightarrow 0$, the
model is indeed in the SOC class; for small but finite $\nu$ the
system is at the edge of SOC and the critical behaviors are still
largely observed; finally, as $\nu$ gets large enough, Gaussian
properties are found. Thus, in the small $\nu$ regime, our model
reproduces the general properties of existing catalogs and can help
interpreting them, while in the large $\nu$ regime it foresees a
different class of behavior. Nevertheless, for the sake of clarity,
we have considered the simple case where the rates $\nu$ and $C$ are
uniform. Thus, the distributions of figure \ref{scaling} may be not
directly comparable to landslide inventories, which gather events
with non-uniform driving rates.

Pictures of a typical ``avalanche'' in the different regimes
discussed above are plotted in figure \ref{map} (upper panels) with
the corresponding values of the factor of safety, $FS_i$, on the
model grid after the avalanche (lower panels). The snapshots in the
upper panels, taken in two systems driven at different rates (left
$\nu=5\cdot 10^{-3}$, right $\nu=5\cdot 10^{-2}$), show a typical
event with size $s=230$ (such a value is chosen because it has
approximately the same probability in the two cases, see figure
\ref{scaling}):
the system on the left is in the power-law regime; 
the one on the right is in the non power-law regime. 
The difference of the avalanche geometry in the two cases is
impressive. Domino effects are crucial to determine a
``catastrophic'' event when the system is governed by a power-law
statistics, where a huge compact landslide is present
\citep{Pietronero_91} with a typical size of the order of the system
size (left-upper panel). Conversely, large events are expected at
higher $\nu$ (where indeed the average size $\langle s\rangle$ is
much larger than $230$), but in such a regime a typical event with
$s=230$ is made of many tiny unconnected avalanches (summing up to
$s=230$).

Interestingly, even though the $P(s)$ is very different in the two
cases, the probability distribution, $P(FS)$,
of the spatial values of $FS$ on the grid has a
similar Gaussian shape, with comparable averages
$\langle FS\rangle$ and fluctuations $\langle \Delta FS^2\rangle$
($\langle FS \rangle=2.20$ and $\langle \Delta FS^2\rangle=0.16$
for $\nu=5\cdot 10^{-3}$; $\langle FS \rangle=2.53$ and
$\langle \Delta FS^2\rangle=0.06$ for $\nu=5\cdot 10^{-2}$)
laying far above the instability threshold $FS=1$.
Thus, a measure of just an average safety factor on the
investigated area could provide only very partial information about
the statistics governing landslide events.

The origin of the striking difference of the $P(s)$ in the two
considered cases traces back to the relative extension of spatial
correlations of the factor of safety, $FS$, which is derived from the
correlation function:
\begin{equation}
C(\vec z)={\langle FS(\vec r)FS(\vec r+\vec z)\rangle-\langle FS(\vec
  r)\rangle^2 \over \langle FS(\vec r)^2\rangle-\langle FS(\vec r)\rangle^2}
\end{equation}
where $FS(\vec x)$ is $FS$ at position $\vec x$ (here we take $\vec
z$ along the direction of the slope) and the average is over the
system sites. As it is well-known \citep{Jensen_98}, we find that
$C(\vec z)\propto \exp(-z/\xi)$, where $\xi$ is the spatial
correlation length of $FS$. The value of $FS_i$ on site $i$ is shown
in the lower panels of figure \ref{map}, in gray scale, for the same
cases pictured in the upper panels:
patterns of large correlated areas (regions with similar values of
$FS_i$, i.e., the same color) are apparent in the left bottom panel
and, in practice, absent in the right one. In the power-law regime
(e.g., $\nu=5\cdot 10^{-3}$), the correlation length, $\xi$, is of
the order of the system size; thus, even a very small perturbation
(say, a drop of water) at one single point can trigger huge system
responses. Instead, in the non power law regime (e.g., $\nu=5\cdot
10^{-2}$) large-scale correlations are absent; here, large events
trivially occur just because the strong external driving rate makes
likely that many cells simultaneously approach the instability
threshold. The detection of patterns of correlated domains (i.e.,
the size of $\xi$) in investigated areas results, thus, to be a
crucial tool to identify the response of the system to
perturbations, i.e., for hazard assessment.

{\bf Conclusions } To summarize, we have investigated a continuously
driven anisotropic cellular automaton model for the characterization
of landslide size distributions. The model may help in interpreting
the general behaviors observed in real systems. In particular, we
have found that different values of the driving rate give rise to
different statistical distributions of events. The determination of
correlated domains in the factor of safety becomes crucial for
landslide classification and, consequently, for hazard assessment.

%
%

\begin{acknowledgments}
Work supported by MIUR-PRIN '04, CRdC-AMRA.
\end{acknowledgments}

%
%
%
%
%
%
%
%



%
%
%

\begin{figure}
\caption{The probability density distribution, $P(s)$, of avalanches
of size $s$ is plotted for the shown values of the time derivative of
the inverse factor of safety, $\nu$
(model size $L^2=64\times64$, conservation level
$C=0.4$, anisotropy coefficients $f_u/f_d=2/3$ and $f_l/f_d=5/6$).
The power law $P(s)\sim s^{-\alpha}$ found in the limit
$\nu\rightarrow 0$ is partially preserved by increasing $\nu$ up to a
point where a bell shaped behavior is clearly observed.
}
\label{scaling}
\end{figure}

\begin{figure}
\caption{The probability distribution, $P(s)$, of avalanches of size
$s$ is plotted for the shown values of level of conservation, $C$
($\nu=5\cdot 10^{-3}$, other param.s as in figure \ref{scaling}).
The {\bf inset} shows the exponent $\alpha$ (confidence interval
$95\%$) of the power-law fit as a function of $C$.
}\label{scaling_C}
\end{figure}

\begin{figure}
 \caption{{\bf Top panels}: Two snapshots of a typical landslide event of size
$s=230$ on our $64\times 64$ grid, in two cases with a driving rate $\nu
=5\cdot 10^{-3}$ ({\em left figure}) and $\nu =5\cdot 10^{-2}$
({\em right figure}). The cells marked in black are those which reached
the instability threshold.
{\bf Bottom panels}: The pictures plot the local value of the
factor of safety, FS, corresponding to the stable configurations
reached after the avalanches shown in the upper panels. The FS values
have been associated to ten
levels of color from white to black, in order to measure the
distance of a cell from its instability condition: the darker the
color, the farther is the cell from the instability threshold. In
the panels it is possible to recognize as dark areas the
avalanches shown in the corresponding upper grids. In particular,
dark areas are related to previous landslide events, as
the lighter areas indicate regions of future events.
In the {\em left figure} large correlated regions (compact areas with
same color) are observed, whereas their size is small in the
{\em right figure}.
}\label{map}
\end{figure}

%
%

\end{article}

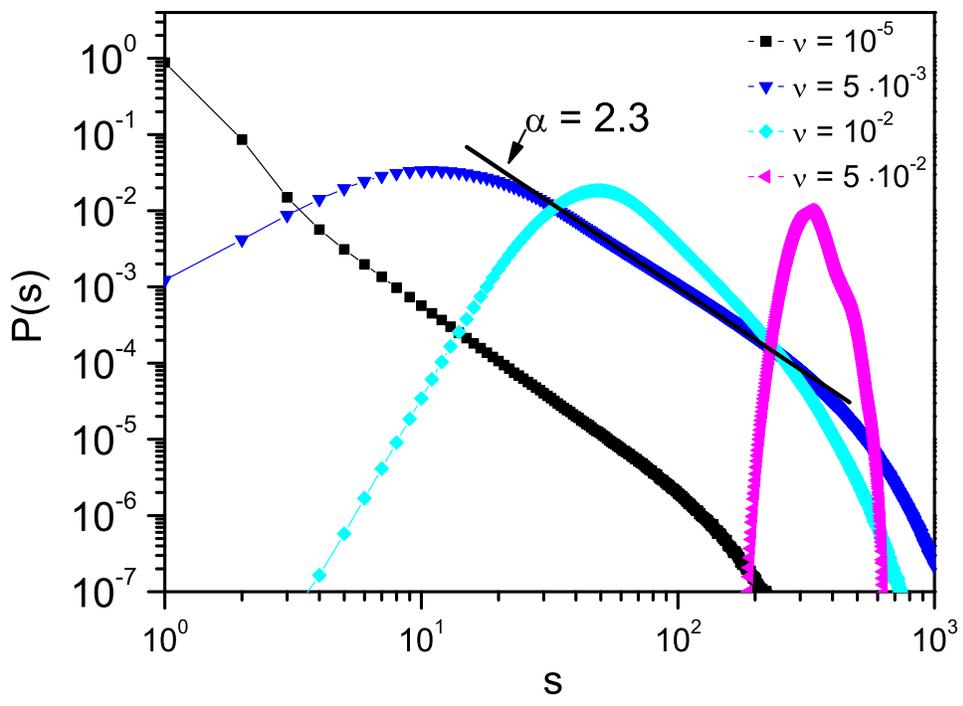
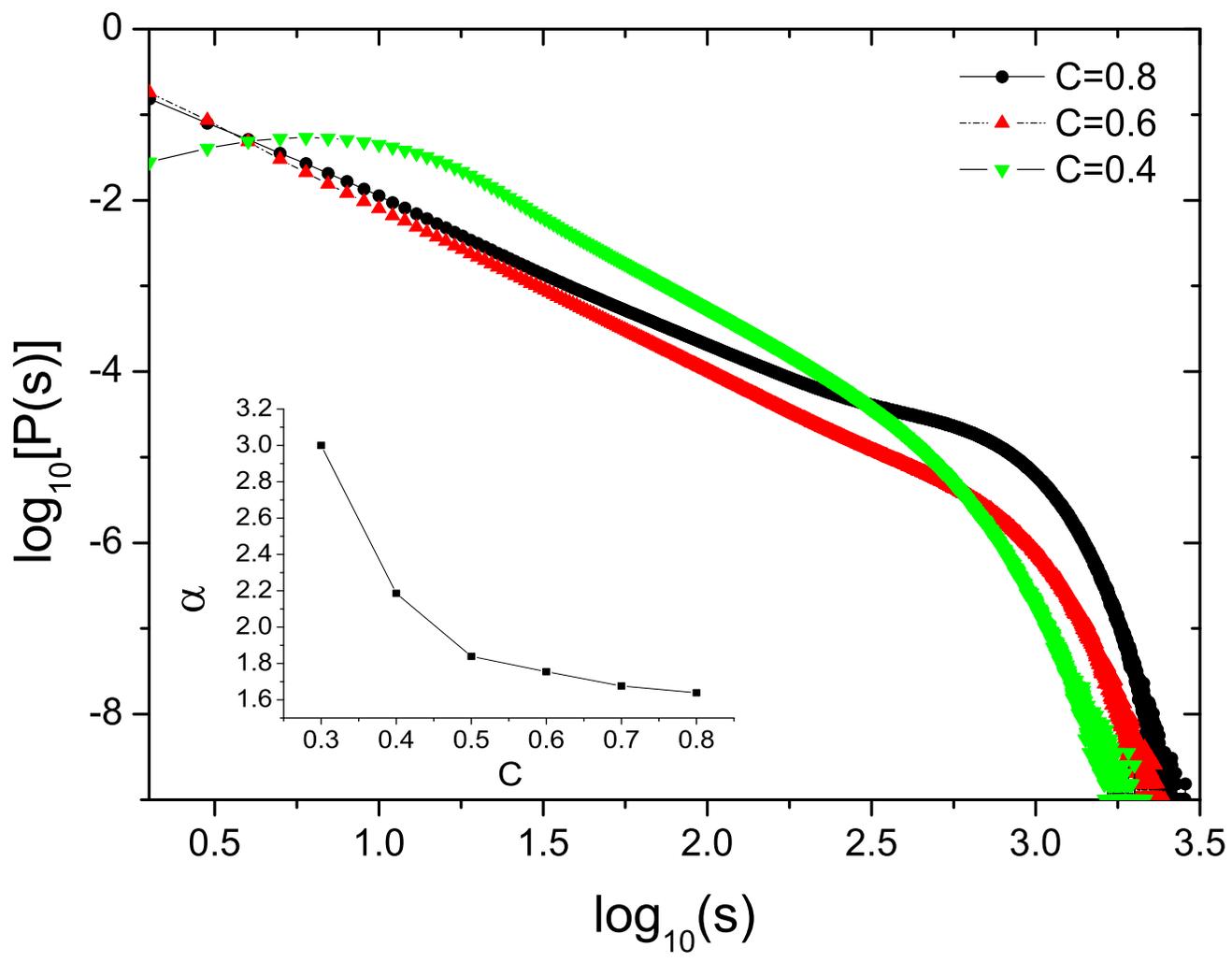
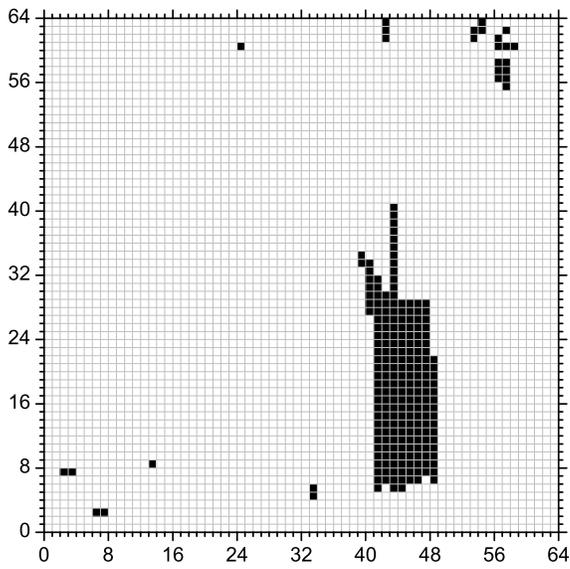
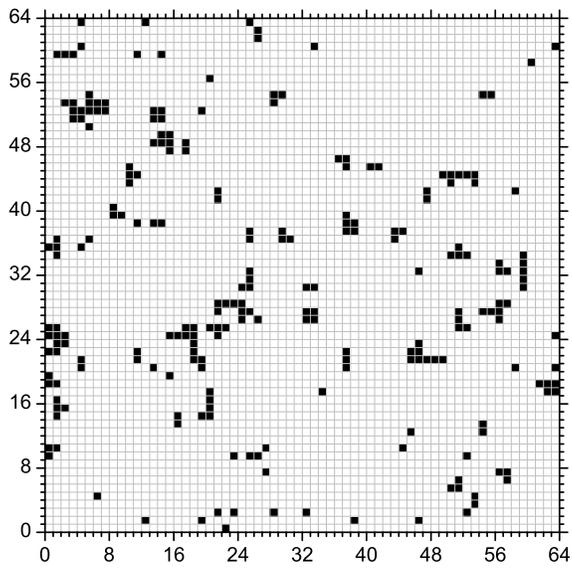
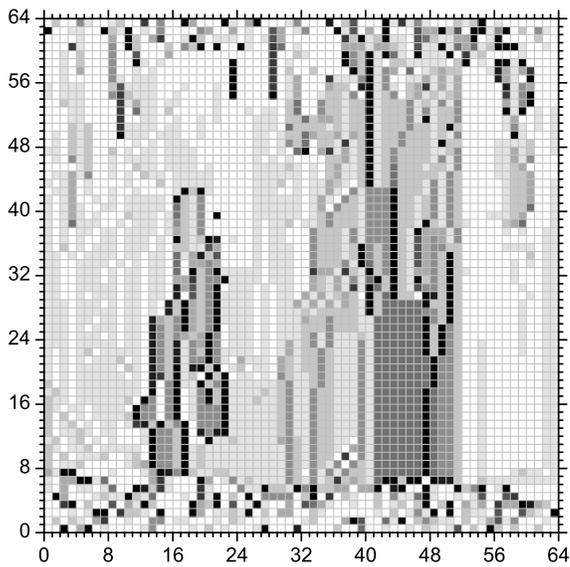
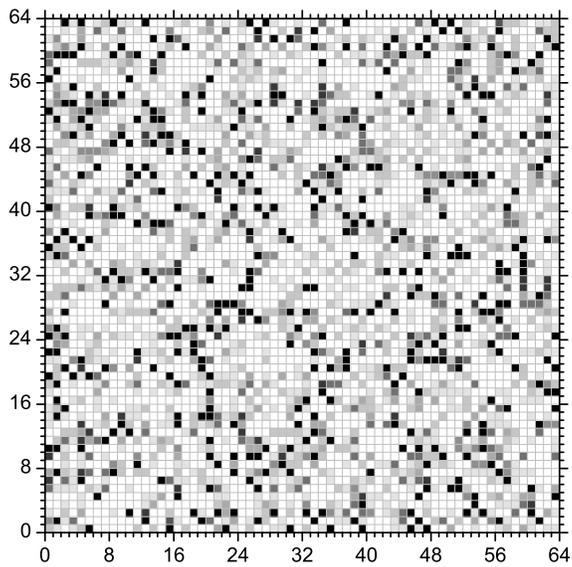
\end{document}